\documentstyle[epsf]{cupconf}

\def\subr #1{_{{\rm #1}}}

\def\etal{{\it et al.}}
\def\cf{{\it cf.}}
\font\ninerm=cmr9
\font\ninei=cmti9
\def\cptn #1{{\noindent{\baselineskip 8pt \ninerm #1}}}

\def\eps@scaling{.95}
\def\epsscale#1{\gdef\eps@scaling{#1}}

\def\plotone#1{\centering \leavevmode
    \epsfxsize=\eps@scaling\columnwidth \epsfbox{#1}}

\begin{document}
\input psfig.tex

\title[Globular-Cluster Color-Magnitude Diagrams]{HST Observations of \\
High-Density Globular Clusters\footnote{Based on observations with the
NASA/ESA {\it Hubble Space Telescope}, obtained at the Space Telescope
Science Institute, which is operated by AURA, Inc., under NASA contract
NAS5-26555}}

\author[G.\ Piotto {\it et al.}]{%
 G.\ns P\ls I\ls O\ls T\ls T\ls O$^1$,\ns
 C.\ns S\ls O\ls S\ls I\ls N$^2$,\ns
 I.\ns R.\ns K\ls I\ls N\ls G$^2$,\\
 S.\ns G.\ns D\ls J\ls O\ls R\ls G\ls O\ls V\ls S\ls K\ls I$^3$,\ns
 R.\ns M.\ns R\ls I\ls C\ls H$^4$,\ns
 B.\ns D\ls O\ls R\ls M\ls A\ls N$^5$,\\
 A.\ns R\ls E\ls N\ls Z\ls I\ls N\ls I$^6$,\ns
 S.\ns P\ls H\ls I\ls N\ls N\ls E\ls Y$^3$,\ns
 \and\  J.\ns L\ls I\ls E\ls B\ls E\ls R\ls T$^7$}

\affiliation{
 $^1$Dipartimento di Astronomia, Universit\`a di Padova, Vicolo
 dell' Osservatorio 5, I-35122 Padova, Italy\\[\affilskip]
 $^2$Astronomy Dept., University of California, Berkeley, CA 92720-3411,
 USA\\[\affilskip]
 $^3$Astronomy Dept., MS 105-24, California Institute of 
 Technology, Pasadena, CA 91125, USA\\[\affilskip]
 $^4$Astronomy Dept., Columbia University, 538 W.\ 120th St., Box 43
 Pupin, New York, NY 10027, USA\\[\affilskip]
 $^5$Laboratory for Astronomy and Solar Physics, Code 681, NASA Goddard
 Space Flight Center, Greenbelt, MD 20771, USA\\[\affilskip]
 $^6$Dipartimento di Astronomia, Universit\`a di Bologna, Cp.\ 596, 
 I-40100 Bologna, Italy\\[\affilskip]
 $^7$Steward Observatory, University of Arizona, Tucson, AZ 85721, 
 USA}
\makepptitle

\begin{abstract}
\noindent
In this paper we present preliminary results from our HST 
project aimed at exploring the connection between stellar
dynamics and stellar evolution in the cores of high-density globular 
clusters. 
\end{abstract}

\firstsection
\section{Introduction} 

Until a few years ago Galactic globular clusters (GC) were regarded as
ideal laboratories for testing stellar evolution theories and for
studying stellar dynamics in {\it simple} stellar systems, and as
important relics of the formation of the Galaxy.
For many years most of the
research proceeded as if the problems in the single fields mentioned
above could be understood and solved independently of each other.

Now, however, each branch of GC studies is realizing that further 
progress depends
on viewing each cluster as a kind of {\it ecosystem} of interrelated
species. 
We have a growing body of observational evidence that
dynamical interactions among stars in high-density clusters can modify
their stellar content. Color gradients have been observed from the
ground in the central regions of some post-core-collapse (PCC) or
high-concentration clusters (\cf\ Djorgovski \& Piotto 1993 for a
review and references). The gradients are
always in the sense of a bluer center, and extend even to the far-UV
wavelengths. The gradients reflect radial
changes in the stellar population. In addition, 
Fusi Pecci, Ferraro, \& Cacciari (1993) have shown that the length of
the horizontal branch (HB), and the presence and the extent of blue
tails in particular, are correlated with the cluster density and
concentration, in the sense of more concentrated or denser clusters
having bluer and longer HB morphologies.  The theoretical
understanding of these phenomena remains unclear, though it is very
likely that binaries and stellar interactions are involved in
modifying stars located on the evolved branches of the color--magnitude
diagram (CMD).

The exceptional resolving power of HST is of fundamental importance in
this kind of study, as it allows observing faint stars in the
center, down to the main sequence. As an example, in Figure~1 we
compare one of the best ground-based images on which our previous
investigation (Djorgovski \& Piotto 1993) was based with the
corresponding HST frame.

\medskip
\hbox{
\vbox{\hsize 2.5 truein
\psfig{figure=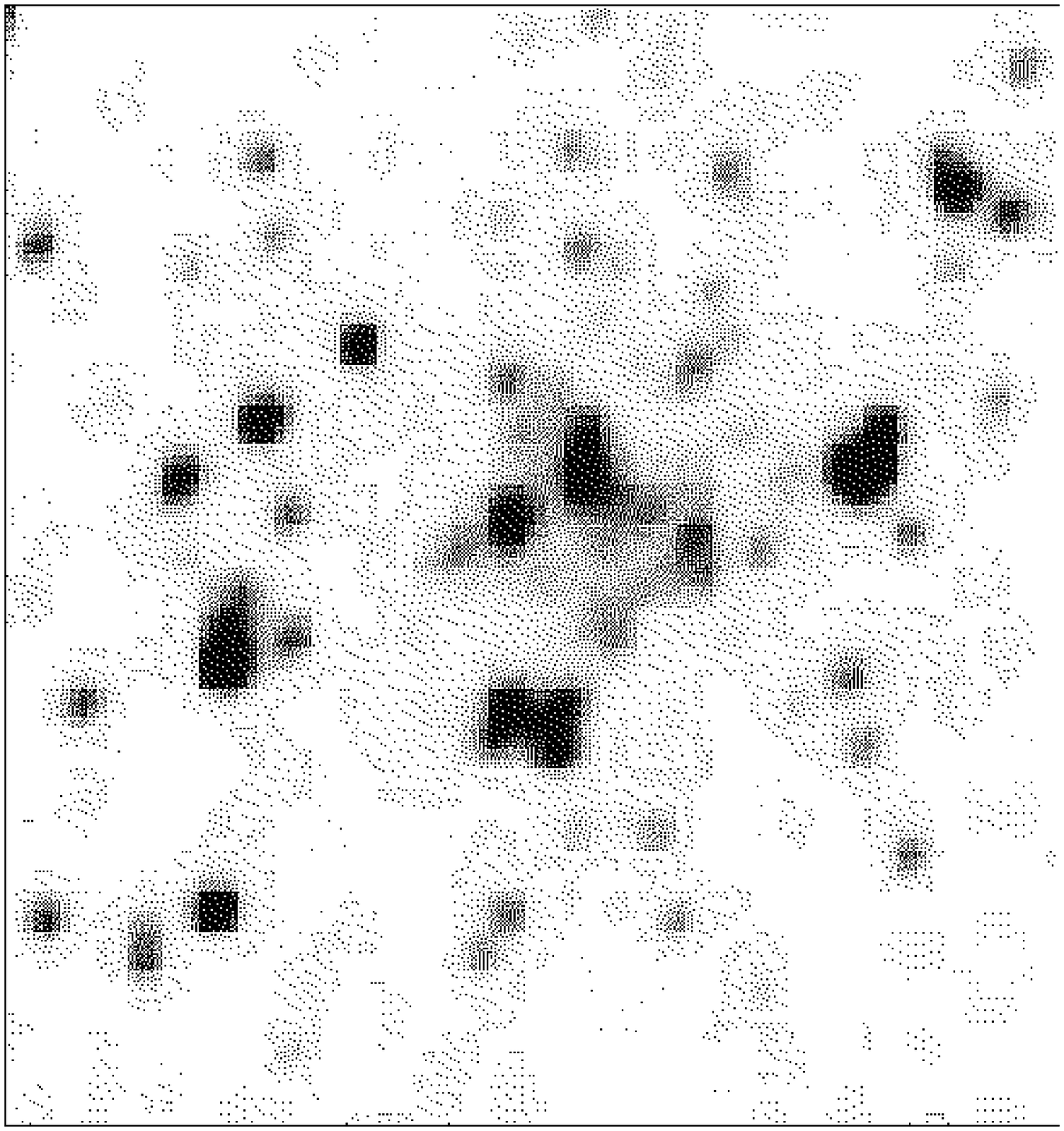,height=2.6truein,width=2.6truein}
}
\vbox{\hsize 2.5 truein
\psfig{figure=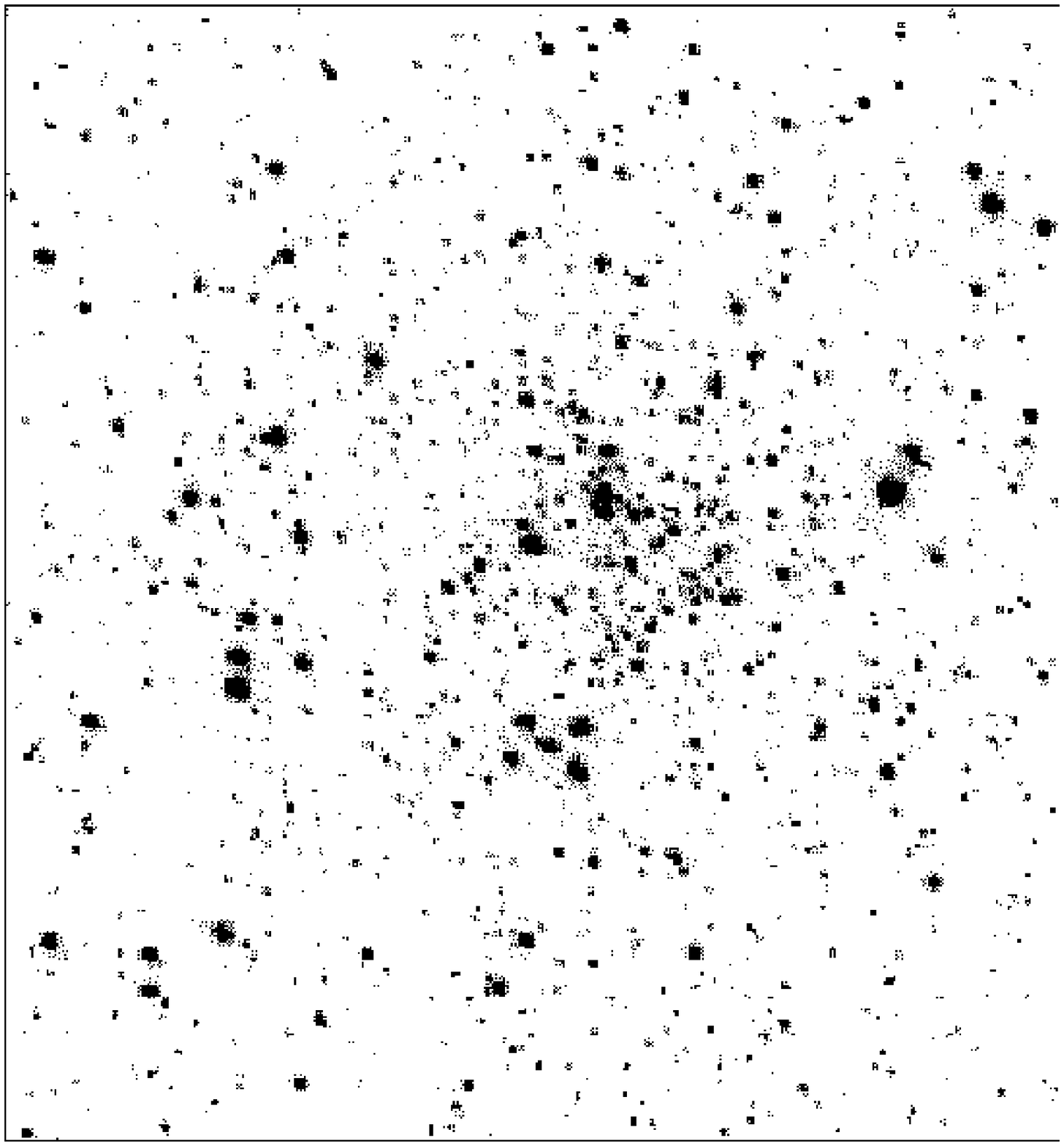,height=2.6truein,width=2.6truein}
}
}
\cptn{Figure 1. The {\it left panel} shows a ground-based {\it V\/}-band 
image of the 
inner $21\times21$ arcsec$^2$ of the GC M30, taken at the ESO NTT telescope
(seeing FWHM = 0.6 arcsec). The {\it right panel} shows the same field
from the PC frame of the WFPC2 camera in the F555W band.}

\bigskip
In this paper we present some preliminary results from our HST project
specifically devised to explore the connection between 
stellar dynamics and stellar evolution by investigating the GC star
population in the very inner regions of 10 high-density clusters.

\section{Observations and data reduction}
The central regions of 10 GCs were observed with the WFPC2 camera on
HST in the F218W, F439W, and F555W bands, during Cycle 6.  The targets were
chosen on the basis of either their high density and/or concentration
or because of the strong UV flux of unknown origin detected by IUE. In
our analysis, we used also the archive data (from an HST program
by Yanny \etal) on 3 additional PCC clusters: NGC~6624, NGC~7078,
and NGC~7099.  
Our cluster sample spans a factor of 100 in metallicity, covering almost the
entire GC metallicity range.  In this paper we focus on the F439W
and F555W images.

The stellar photometry has been obtained with DAOPHOT, 
setting the parameters as discussed in Cool \& King (1995). The F439W
and F555W instrumental magnitudes have been transformed into 
standard $B$ and $V$ magnitudes following Holtzman \etal\ (1995). 
There is a general agreement in the photometric zero point with the
previous ground-based investigations, with the noticeable exception of
NGC 2808, which seems $\sim0.1$ mag fainter in $V$ and redder than in 
Ferraro \etal\ (1990).

\section{The color--magnitude diagrams}
The CMDs for the 13 clusters are shown in the following paper by Sosin \etal.

From a minimum of $\sim 3000$ (NGC~6652) to more than 27000
(NGC~7078) stars have been identified in each cluster, 
from $\sim 2$ magnitudes below the TO to the tip of the giant
branch (GB).  Such a large sample of stars is of particular importance to
check the stellar clock (Renzini \& Fusi Pecci 1988), and a
comparison with the theoretical luminosity functions (LF) is
in progress. 

All the evolved branches of the CMD are clearly identifiable in the 13
clusters. In particular, we note that all the 13 clusters show a 
well-defined blue straggler (BS) sequence, extending from the TO to about
2.5 magnitudes brighter. In only a couple of cases (NGC 6441 and 
NGC 6522) does the strong field contamination make it difficult to extract the 
true  cluster BS population.
A population of supra-HB stars is visible in at least 7 clusters:
NGC 1851, 1904, 2808, 6388, 6441, 6522, and 7078.

The most exciting features in the present set of CMDs concern the HB.  
Six clusters (NGC~1904, 2808, 6388, 6441, 6552, 
7078), plus probably also NGC~362, show extended blue HB tails. In all cases
the blue tail seems to extend down to the photometric limit of our
data. In particular, the HB of NGC~2808 is densely populated
down to at least two magnitudes below the TO. There are
at least two unexpected and, at the moment, inexplicable findings:

\begin{itemize}

\item the blue horizontal branches, with extended blue tails, in the two 
metal-rich clusters NGC 6388 ([Fe/H] $=-0.60$) and NGC 6441 ([Fe/H] $=-0.53$);

\item the clumpy nature of the extended blue HB tail of NGC~2808.

\end{itemize}

\subsection{NGC 6388 and NGC 6441}
A more detailed discussion of NGC 6388 and NGC 6441 will appear in 
Rich \etal\ (1997).
\medskip
\hbox{
\vbox{\hsize 2.5 truein
\psfig{figure=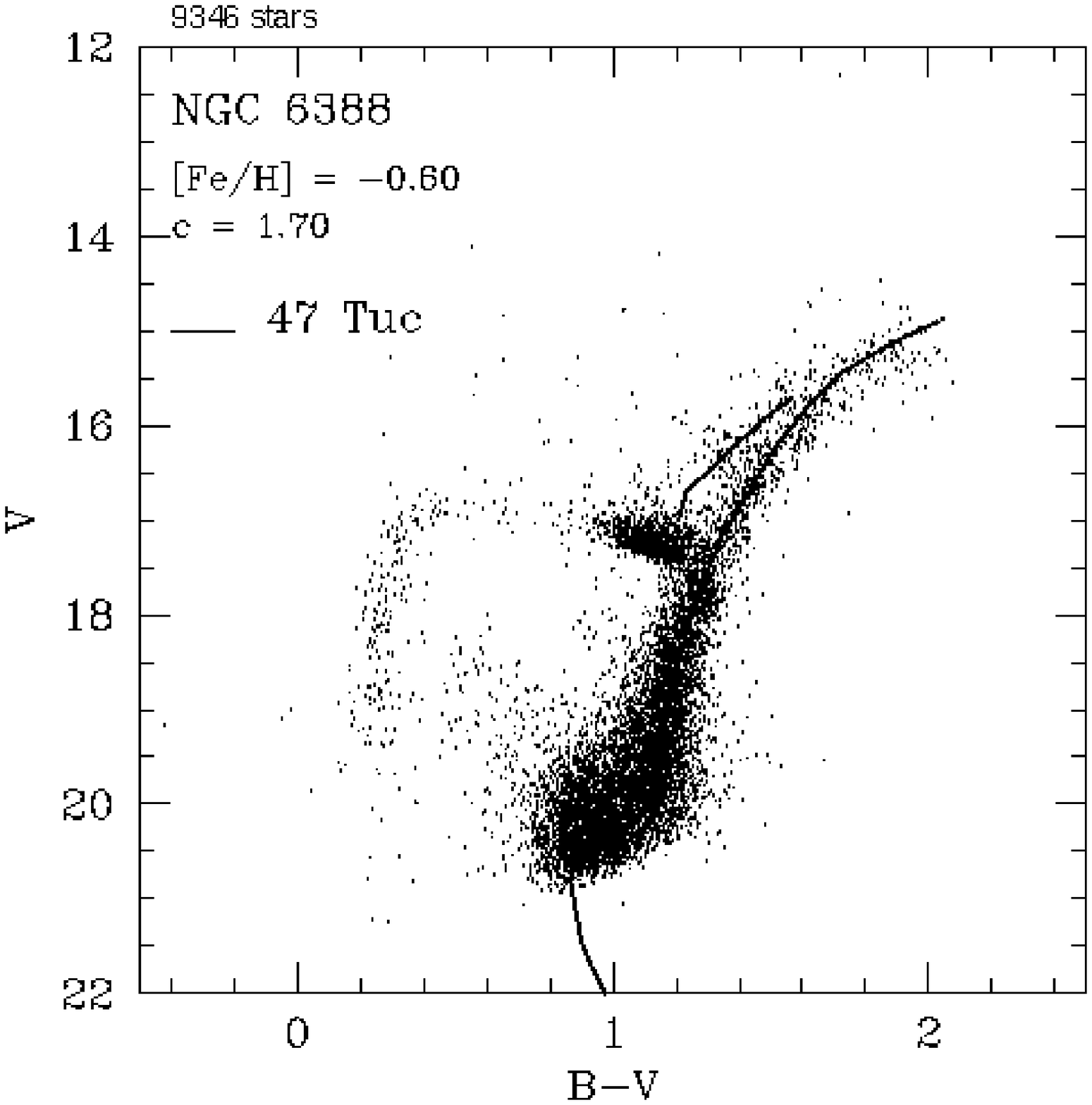,height=2.6truein,width=2.6truein}
}
\vbox{\hsize 2.5 truein
\psfig{figure=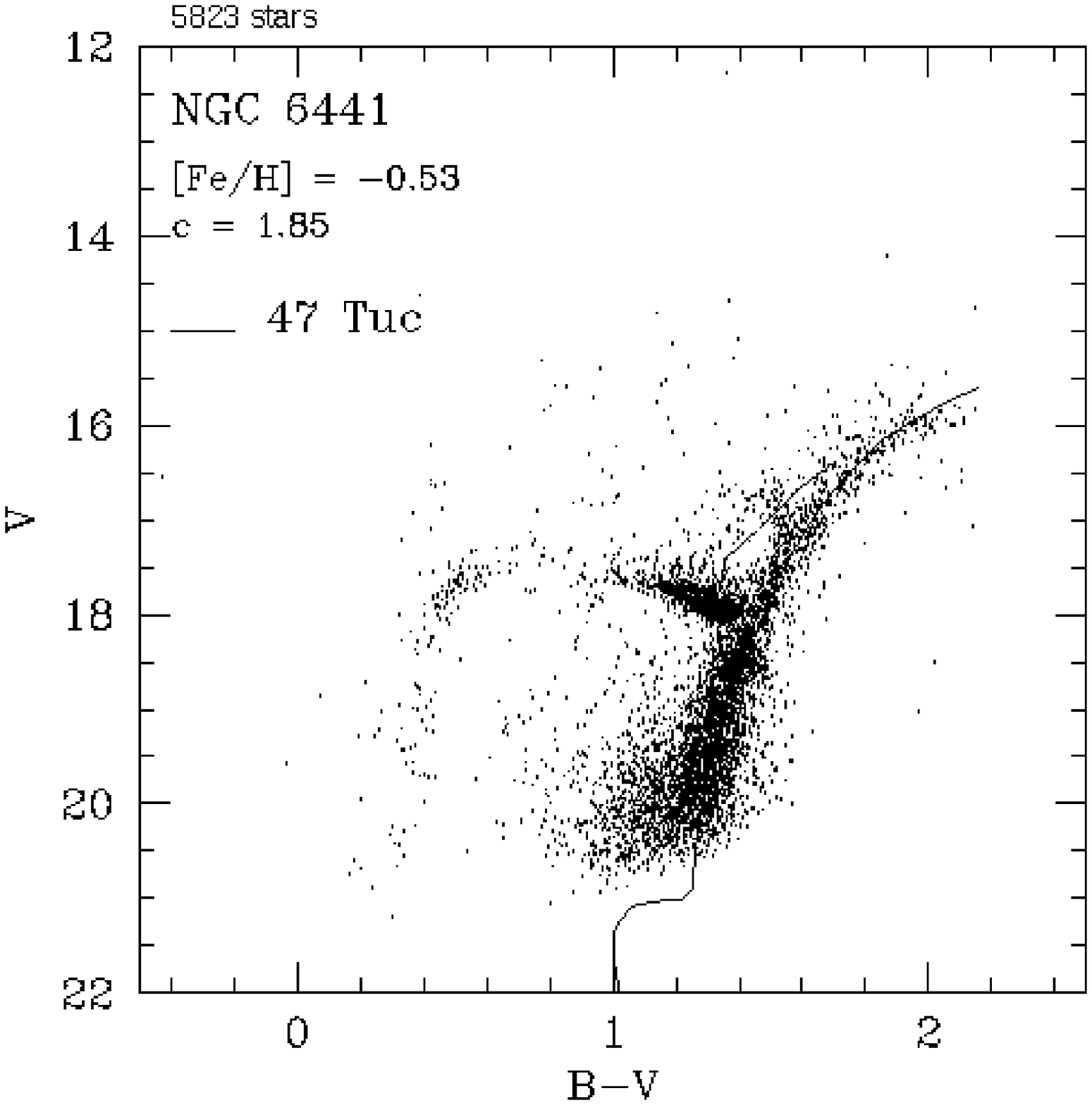,height=2.6truein,width=2.6truein}
}
}
\cptn{Figure 2. (left) The ($B$, $B-V$) CMD of NGC 6388 is compared with the
fiducial sequence of 47 Tuc.  The slopes of the giant branches of the two 
clusters are similar (the NGC 6388 giant branch is marginally flatter),
suggesting similar metal content.
(right) As in the left panel, but for NGC 6441. Again, the metallicity of
NGC 6441 cannot differ much from that of 47 Tuc. 
The color extent of the red HBs of NGC 6388 and NGC 6441 is greater than
that of the HB of 47 Tuc.}

\bigskip
Regarding the average metal content of NGC 6388 and NGC 6441 there are
very few doubts.
Figure~2 reproduces their CMDs. For comparison,
we have over-plotted the fiducial points of the CMD of 47~Tuc
([Fe/H] $=-0.71$). This comparison clearly shows that the giant branches of
NGC~6388 and NGC~6441 have similar slopes to that of---or are 
marginally flatter 
than---the giant branch of 47 Tuc. This means that, on average, 
they have the same---or slightly higher---metallicity than 47~Tuc. 
On the basis of their metal content we would expect to see only a red 
stub of the HB. Instead, we see also a prolonged HB tail, which extends to the
limit of our photometry in both of them, with a clear presence of at least
one and possibly two gaps, which remind us of the gaps in NGC 2808. No standard
evolutionary model can reproduce these blue tails and the red HB at the 
same time. We note here that in view of the presence of both the (unexpected)
blue tail and of the red HB stub, NGC 6388 and NGC 6441 might be regarded as
two further examples of clusters with a bimodal HB (like NGC 1851 and 
NGC 2808), or better, as the most metal rich among the known bimodal-HB
GCs. Whatever the origin of the blue tail is, NGC 6388 and 
NGC 6441 tell us that a truly metal-rich old population can also create 
hot blue stars, and this result might be related to the ultraviolet
flux increase towards shorter wavelengths discovered in elliptical
galaxies (Bertola \etal\ 1980).

A close inspection of the CMDs in
Figure 2 (\cf\ also Figure 1 in Sosin \etal)
shows that
the red HB is peculiar, as is in general the entire CMD. 
Indeed, we can compare the
CMDs of NGC 6388 and NGC 6441 with the CMDs
of the other metal-rich clusters in the present sample, i.e., 
47 Tuc, NGC 5927, and NGC 6624. In these three clusters, the red HB (RHB)
is very well defined, well confined to a
restricted color interval, almost parallel to the color axis, and very well
separated from the red GB (RGB). 
NGC~6388 and NGC~6441 have a completely different RHB,
spread out in a larger color interval, inclined with respect to the 
color axis, merging with the RGB on the red side.
These features might be thought to be due to differential reddening or 
to crowding effects.
We are presently running artificial-star tests, but it is unlikely that
internal photometric errors can be the explanation, as the effect is perfectly 
visible in a similar way also in the CMDs from the WF3 and WF4 chips alone, 
which are much less crowded.
There might be some differential reddening, though 
the average reddening is not very high ($<0.4$ mag)
and the covered field is very small, particularly with the PC camera, while
the effect is visible in every single WFPC2 chip.
A direct comparison of the CMDs in different regions of NGC 6388 shows that
there is no average zero-point difference at scales larger than about 8 
arcsec. So if there is some differential reddening, it must be present
at scales smaller than 8 arcsec, still leaving the average reddening the 
same at large scales in the chip area of about $160\times160$ arcsec$^2$.
On the contrary, there might be some differential reddening in NGC 6441,
at the level of a few hundredths of a magnitude.
Differential reddening might be a possibility, though it is does not seem
likely, at least for NGC 6388.

There is another possibility, which might be an interesting working 
hypothesis.
At a first glance, the CMDs of NGC 6388 and NGC 6441 strongly recall the
CMD of $\omega$ Cen, apart from the RHB, raising the suspicion
that also in the former two clusters there might be some spread in 
metallicity.
At least qualitatively, a spread in metallicity might be 
an appealing explanation for the anomalous HB and for the spread in
the RGB. Of course, this is just a 
possibility, suggesting that it might be of some interest to study the
metal content of a few single stars in these clusters.

Another working hypothesis is that there might be two stellar populations in 
the cluster, with two different ages. 
This possibility has been excluded for all the other
known bimodal GCs, and it seems unlikely that it can work
for NGC 6388 and NGC 6441. Unfortunately, the present 
material does not allow testing this hypothesis. 

There is a further possible explanation for the BHB tail in NGC 6388 
and NGC 6441: it might well be possible that their BHBs originate from the 
same phenomenon (whatever it is) of core dynamics that is responsible
for the bluer 
and longer HB tails in the clusters with denser cores (Fusi Pecci \etal\
1993, Buonanno \etal\ 1997). Again, NGC 6388 and NGC 6441 would be the
most metal-rich clusters in which the phenomenon has been discovered.
Tidal stripping of red-giant envelopes during 
close stellar encounters has been suggested as a possible cause 
(see, however, Djorgovski \etal\ 1991). 
Indeed, Rich \etal\ (1997) have calculated 
that NGC 6388 and NGC 6441 have among the highest collision rates 
for any globular cluster in the Galaxy. It is therefore possible that these
high collision rates are responsible for the extended BHB. 
But this scenario presents a problem: if tidal collisions are 
responsible for the production of the hot HB stars, we would expect them
to have a
different radial distribution compared to other evolved stars in the cluster.
However, we find that blue and red HB stars, subgiants, red giants, and
asymptotic GB stars all have the same radial distribution (in these 
and in all the other 11 clusters), and
this might be a serious problem for the tidal-stripping model.

\subsection{NGC 2808}
A detailed discussion of the results on NGC 2808 will be presented by Sosin
\etal\ (1997).

\medskip
\centerline{\psfig{figure=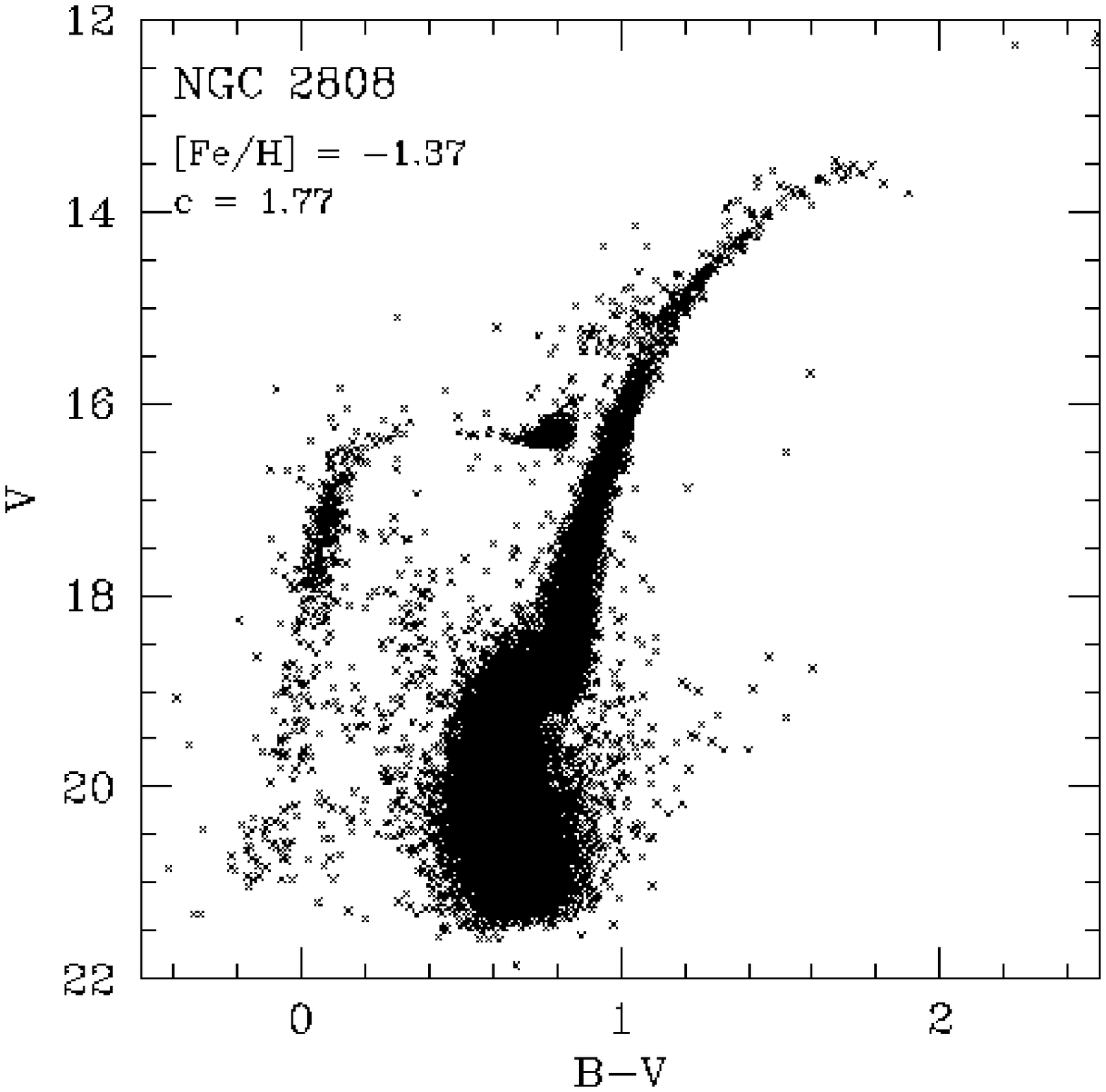,width=4truein}}
\cptn{Figure 3. The $V$ vs. $(B-V)$ CMD for 27286 stars in NGC 2808. Note 
that the HB population is divided into four groups, separated by three gaps.}

\bigskip
The ($V$, $B-V$) CMD of NGC 2808 is presented in Figure 3.
This cluster was already known
to have a bimodal HB with a blue and red HB more or less 
evenly populated. What is new here is that the long blue tail of the HB 
extends down to $V=21$, at the limit of our photometry. The blue tail is not
evenly populated: there are two significant gaps 
(\cf\ also Figure~2 of Sosin \etal, in the following paper)
at $V=18$ and at $V=20$
and colors ${\rm F218W}-B=-1.3$ and $-2.0$\footnote{Note that while the 
F439W and F555W magnitudes have been
converted to Johnson {\ninei B} and {\ninei V} magnitudes, the 
F218W magnitudes have been
calibrated to the STMAG instrumental system, using the zero points in
the tables of Holtzman {\ninei et} {\ninei al}.\ (1995).}, 
corresponding to effective temperatures $\log T\subr{eff}$ = 4.23 and 4.40, 
respectively. As shown in Sosin \etal\ (1997), the
two gaps in the blue tail correspond to masses near 0.54 and 0.495 $m_\odot$,
and the widths of the depleted regions in mass are very narrow, 
$\sim0.01$ $m_\odot$.
To these two gaps we have to add also a third gap, between the red and blue
HB. 
As far as the discontinuities in the HB distribution are concerned,
NGC 2808 is not unique. If we look at the 13 CMDs in the present
paper, we will see
gaps which recall the gaps in the blue tail of NGC 2808 (but are not
necessarily similar or of similar origin) also in NGC 1904, 
NGC 6441, NGC 7078, and perhaps NGC 6388. Moreover, two significant 
gaps at more or less the same temperature as the gaps in NGC 2808 are also
present in the blue HB of high-latitude halo subdwarfs (Newell 1973). So gaps
might be not an exceptional feature in the blue HB. 

It is not clear what the origin of these gaps is. In any case, either the
BHB stars have a particular set of properties which make them 
land in well-defined regions of the HB after the helium flash, or the stars
which happen to fall on the gap regions rapidly evolve off them. While this
last possibility might (at least partially) explain the intermediate gap
(close to the mass where the HB evolution changes from predominantly redward 
to blueward), no obvious reason can be found for the other two gaps ({\it cf.}
Sosin \etal\ 1997).

As in the case of NGC 6388 and NGC 6441, at the moment we have no 
satisfactory explanation for the multi-modality of the HB of NGC 2808.
It might be due to dynamical effects, mass-loss processes, or a 
combination of factors. For sure, the peculiar HBs of these clusters
strongly remind us how poor is our knowledge of the mass-loss processes,
which are the main cause of the HB morphology, as they set the
final stellar envelope mass, i.e., where the stars should move after
the helium flash.

\section{The blue stragglers}

\medskip
\hbox{
\vbox{\hsize 2.5 truein
\psfig{figure=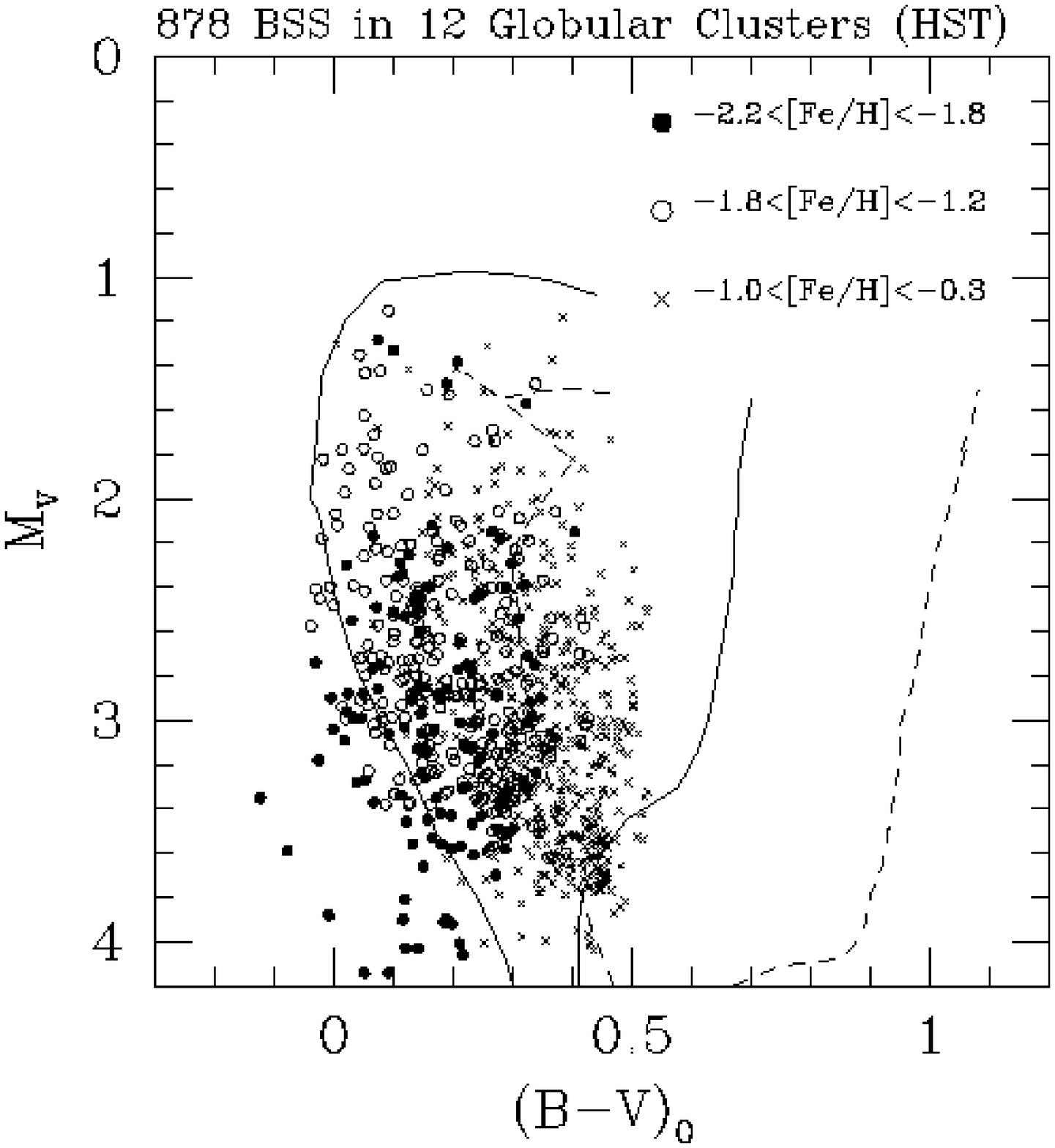,height=2.6truein,width=2.6truein}
}
\vbox{\hsize 2.5 truein
\psfig{figure=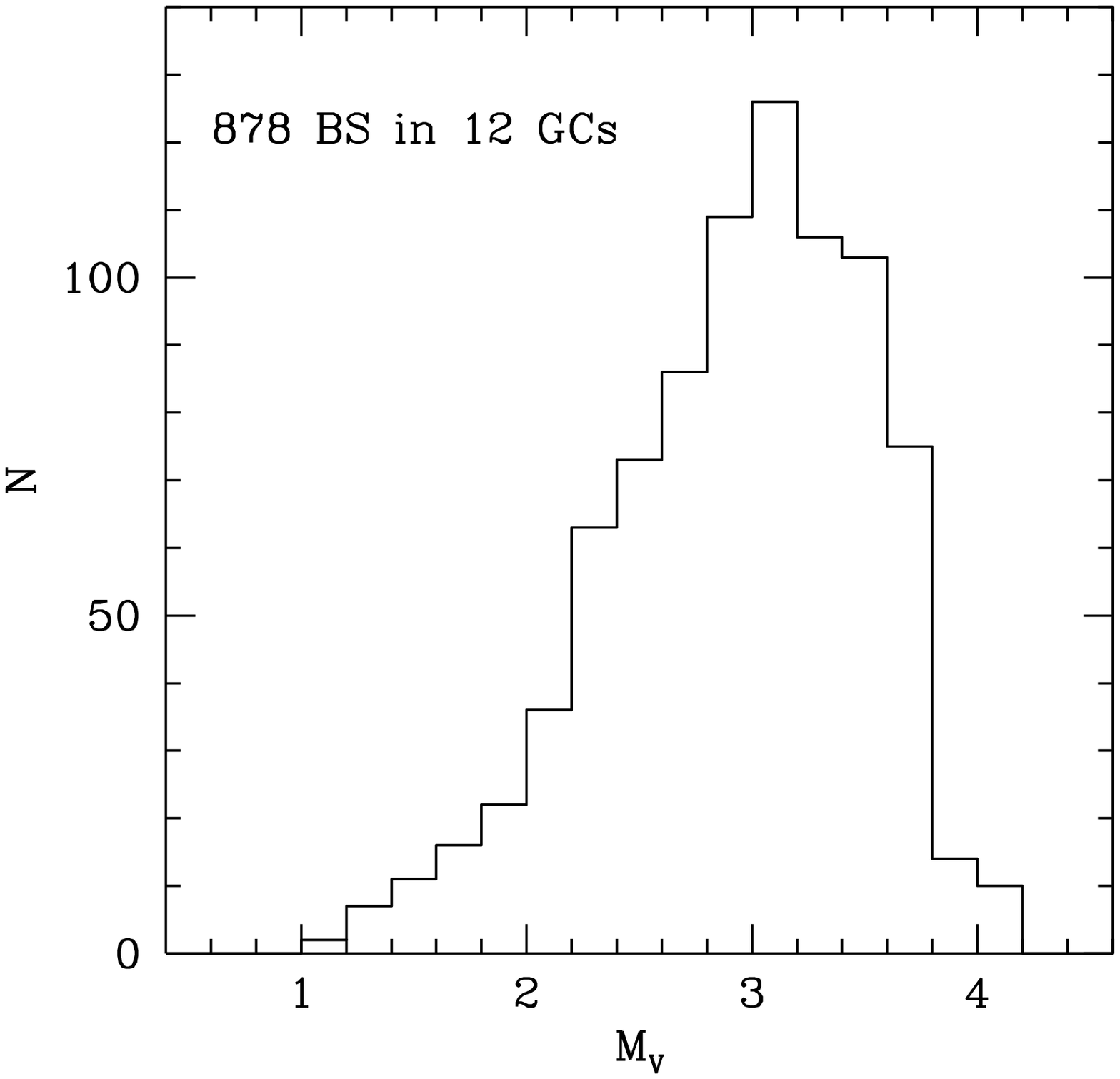,height=2.6truein,width=2.6truein}
}
}
\cptn{Figure 4. The {\ninei left} {\ninei panel} shows The 
{\ninei M$_V$}, {\ninei (B$-$V)}$_0$ CMD for 
the 878 BSs selected from 12 clusters in our sample.  The BSs have
been divided into 3 metallicity groups. The
isochrones for a 1.6 $m_\odot$ and a 0.8 $m_\odot$ star with 
metallicities $Z=10^{-4}$ and $4\times10^{-3}$ are also plotted. 
The {\ninei right} {\ninei panel} shows the {\ninei V\/} luminosity 
function for the same BSs.}

\bigskip
\noindent
As already noticed in the previous Sections, all of the 13 clusters in the
present sample have a population of blue stragglers. We will discuss
the properties of the BSs in a forthcoming paper (Piotto \etal\ 1997).

Here we want to present only a few preliminary results. A sample of 878 
BSs have been extracted from 12 of the 13 clusters in our sample (even
though NGC 6522 appears to have a BS population, the contamination 
by field stars prevented us
from extracting its  BSs). This is the largest sample of BSs in GCs
available so far (\cf\ Fusi Pecci \etal\ 1993 for a comparison). 
Moreover, all the BS magnitudes are in a single, photometrically 
homogeneous system.

Figure 4, {\it left panel}, shows the $M_V$,(B-V)$_0$ CMD for the 878
BSs. Distance moduli and reddenings have been extracted from the
compilation by Djorgovski (1993).  The BSs of Figure 4 have been
divided into three metallicity groups (see labels).  As expected, the 
BSs become bluer and bluer as the metallicity decreases.  In the same
figure we have plotted the isochrones (from Bertelli \etal\ 1994) for
a 1.6 $m_\odot$ star for the two metallicities approximately
corresponding to the extreme metallicities of our sample of GCs (the
full line is for $Z=10^{-4}$ and the dashed line is for
$Z=4\times10^{-3}$, respectively). On the red side are the
isochrones for a 0.8 $m_\odot$ star for the same metallicities.
The {\it right panel} of Figure~4 shows the LF from the 878 BSs.

Here we briefly summarize the main results on the BS population:

\begin{itemize}
\item The brightest BSs have magnitudes as bright as, but not exceeding the 
magnitude of the most metal-poor isochrone for a 1.6 $m_\odot$ star.
In other words, the masses of the BSs in our sample
do not exceed twice the mass of a normal TO star (note that their
masses could be smaller, according to the BS evolutionary models by Baylin 
\& Pinsonneault 1995).

\item There are many BSs redder than the $Z=4\times10^{-3}$ isochrone. This 
might be due both to photometric errors and to the evolution of these stars
off the main sequence.

\item It might be harder to explain the presence of a group of
metal-poor BSs significantly bluer than the $Z=10^{-4}$ isochrones. 
We need to run artificial-star experiments, but it seems unlikely that all
those objects come from photometric errors.

\item In all the 12 clusters, a Kolmogorov--Smirnov test has shown that the
BSs are more concentrated than the subgiants of the same magnitudes,
with a confidence level that is always greater than 99.99\%. This result is of
particular interest because this is the only group of stars whose
radial distribution differs significantly from the others.

\item The BSs closer to the center are marginally bluer and significantly
brighter than the BSs farther from the center.  

\item The LF shows that
there are BSs up to 3 magnitudes brighter than their corresponding
TO. The LF rises steeply up to $M_V\sim3.1$, while the significance of the
apparent decline at fainter magnitude is uncertain, due to the possible bias
in selecting the BSs. No significant gaps can be found in the present LF.

\end{itemize}


\begin{thebibliography}{} 

\bibitem[]{}
Baylin, C.D., \& Pinsonneault, M.H. 1995, ApJ, 439, 705

\bibitem[Bertelli et al. 1994]{br94}
  Bertelli, G., Bressan, A., Chiosi, C., Fagotto, F., Nasi, E., 1994, 
  A\&AS, 106, 275 

\bibitem[Bertola et al. 1994]{b94}
Bertola, F., Capaccioli, M., Holm, A.V., Oke, J.B. 1980, ApJ, 237,
L65

\bibitem[Buonanno \etal\ 1997]{buo97} 
Buonanno, R., Corsi, C.E., Bellazzini, M., Ferraro, F.R., \& Fusi
Pecci, F. 1997, to appear in AJ, February 1997

\bibitem[Cool and King 1995]{coo95} 
Cool, A.C., \& King, I.R. 1995, in {\it Calibrating HST:\ Post
Servicing Mission}, eds.\ A.\ Koratkar \& C.\ Leitherer (Baltimore:\ STScI), 
p.\ 290

\bibitem[]{}
Ferraro, F.R., Clementini, G., Fusi Pecci, F., Buonanno, R., \&
Alcaino, G. 1990, A\&AS, 84, 59

\bibitem[Fusi Pecci \etal\ 1993]{ffp93} 
Fusi Pecci F., Ferraro, F.R., \& Cacciari, C. 1993, in {\it Blue 
Stragglers}, ed.\ R.A.\ Saffer, ASPCS 53 (San Francisco:\ ASP), p.\ 97

\bibitem[Djorgovski 1993]{djo93} 
Djorgovski, S. 1993, in {\it Structure and Dynamics of Globular Clusters},
eds.\ S.G.\ Djorgovski \& G.\ Meylan, ASPCS 50 (San Francisco:\ ASP), p.\ 37

\bibitem[Djorgovski \etal\ 1991]{djoet93} 
Djorgovski, S., Piotto, G., Phinney, E.S., \& Chernoff, D.F. 1991,
ApJ, 372, L4

\bibitem[Djorgovski and Piotto 1993]{djopio93} 
Djorgovski, S., \& Piotto, G. 1993, in {\it Structure and Dynamics of
Globular Clusters}, eds.\ S.G.\ Djorgovski \& G.\ Meylan, ASPCS 50 
(San Francisco:\ ASP), p.\ 203

\bibitem[Holtzman \etal\ 1995]{holtzman}
Holtzman, J.A., Burrows, C.J., Casertano, S., Hester, J.J.,
Trauger, J.T., Watson, A.M.,  Worthey, G. 1995, PASP, 107, 1065

\bibitem[Newell 1973]{new73}
Newell, E.B., 1973, ApJS, 26, 37

\bibitem[]{}
Piotto, G., et al. 1997, in preparation

\bibitem[Renzini and Fusi Pecci 1988]{rf88}
Renzini, A., \& Fusi Pecci, F. 1988, ARAA 26, 199

\bibitem[]{}
Rich, R.M., et al. 1997, in preparation

\bibitem[]{}
Sosin, C., et al. 1997, in preparation

\end{thebibliography}
\end{document}